\documentclass[usenatbib, twocolumn, twocolappendix]{aastex631}

\usepackage{graphicx}	
\usepackage{amsmath}	
\usepackage{ulem,soul}
\usepackage{booktabs, threeparttable}
\usepackage[nice]{nicefrac}
\usepackage{hyperref}
\usepackage{xcolor}
\hypersetup{linkcolor=red,citecolor=blue,filecolor=cyan,urlcolor=magenta}
 
\newcommand{\rmd}{{\mathrm d}}

\newcommand{\vect}[1]{\mathbf{#1}}
\newcommand{\avg}[1]{\langle{#1}\rangle}
\newcommand{\abs}[1]{\left\vert{#1}\right\vert}

\newcommand{\lhmpc}{{h^{-1}\rm Mpc}}

\newcommand{\msun}{{h^{-1}\rm M_{\sun}}}

\shorttitle{Halo mass of ULIRGs at $z\sim 2$}
\shortauthors{Pan, Li, Cheng \& Huang}

\begin{document}

\title{Halo Mass of ULIRGs at Cosmic Noon}

\correspondingauthor{M. Li, J. Pan, J.-S. Huang}
\email{mingli@nao.cas.cn, jpan@nao.cas.cn, jhuang@nao.cas.cn}

\author[0000-0002-4025-7877]{Jun Pan}
\affiliation{Chinese Academy of Sciences South America Center for Astronomy, National Astronomical Observatories, CAS, Beijing, 100101, People's Republic of China}
\affiliation{College of Earth Sciences, Guilin University of Technology, Guilin 541004, People's Republic of China}

\author[0000-0002-1318-4828]{Ming Li}
\affiliation{Key Laboratory for Computational Astrophysics, National Astronomical Observatories, Chinese Academy of Sciences, Beijing 100101, People's Republic of China}

\author[0000-0003-0202-0534]{Cheng Cheng}
\affiliation{Chinese Academy of Sciences South America Center for Astronomy, National Astronomical Observatories, CAS, Beijing, 100101, People's Republic of China}

\author[0000-0001-6511-8745]{Jiasheng Huang}
\affiliation{Chinese Academy of Sciences South America Center for Astronomy, National Astronomical Observatories, CAS, Beijing, 100101, People's Republic of China}

\begin{abstract}
We present a clustering analysis of $\sim 3000$ ultraluminous infrared galaxies (ULIRGs) at $z\sim 2$, uniformly selected by $24\micron$ flux and IRAC colors in the COSMOS and BOOTES fields. 
We measure the angular correlation functions of ULIRGs in both fields and fit them with galaxy clustering models. Linear theory modeling shows that these ULIRGs reside in dark matter halos with characteristic masses of $\log M_{\rm h} /(\msun) = 12.64\pm 0.50$ in COSMOS and $12.72\pm 0.13$ in BOOTES. The halo occupation distribution (HOD) modeling yields occupation-weighted effective halo masses of $\log M_{\rm eff} /(\msun) =12.50^{+0.26}_{-0.28}$ for COSMOS and $12.89^{+0.11}_{-0.12}$ for BOOTES. These host halos are expected to evolve into halos with masses of $\sim 10^{13.6-13.9}\msun$ at $z=0$. The HOD fits allow for a non-negligible satellite contribution to the clustering of galaxies in the BOOTES field, but the satellite fraction  derived in the COSMOS field appears nearly zero.
\end{abstract}

\keywords{Ultraluminous infrared galaxies(1735) --- High-redshift galaxies(734) --- Two-point correlation function(1951) -- Galaxy dark matter halos(1880)}

\section{Introduction}
The epoch at redshift $z\sim 2$, often referred to as ``cosmic noon", marks a critical phase in the history of galaxy formation and evolution. During this period, the cosmic star formation rate density reached its high plateau and subsequently began to decline rapidly, with more than half of the present-day stellar mass in galaxies assembled within $1<z<3$ \citep{DickinsonEtal2003, MadauDickinson2014}. Since galaxies form and evolve within halos of cold dark matter, their properties and evolution histories are closely correlated with the local environment \citep[e.g.][]{Dressler1980, KauffmannEtal2004, ShethEtal2006, PapovichEtal2018} and are regulated by their host halos \citep[see review of][]{WechslerTinker2018}. It is therefore particularly important to constrain the host halos of strongly star-forming galaxies, as these systems are expected to play a key role in the buildup of massive galaxies during cosmic noon. Significant evidence suggests that such vigorous assembly with intensive star formation is often triggered by galaxy mergers \citep[e.g.,][]{SandersEtal1988, ColeEtal2000}. The galaxy clustering analysis is commonly regarded as the standard approach to establishing the galaxy-halo connection, providing direct statistics on the physical quantities of galaxies, such as the mass of their host halos. Unfortunately, the redshift interval $1.4<z<2.5$ is often known as the optical {\em redshift desert} \citep{SteidelEtal2004}, and spectroscopic surveys in this range remain observationally challenging. Moreover, large spectroscopic surveys in other bands are prohibitively expensive \citep[e.g.][]{CoxEtal2023}. Consequently, spectroscopic samples constructed for clustering analysis at $z\sim 2$ typically contain only a few hundred galaxies or even fewer \citep[e.g.][]{BlainEtal2004, StachEtal2021}. Photometric samples of star-forming galaxies, selected through various methods, have therefore become essential for clustering analysis in this redshift regime \citep[e.g.][]{FarrahEtal2006, KampenEtal2023}. 

Among the populations of galaxies dwelling at cosmic noon, we focus on very massive and dusty star-forming galaxies, which are extraordinarily luminous in the infrared band and exhibit star-formation rates (SFR) of $\sim 10^2-10^3M_\sun/yr$ or even higher. These galaxies are recognized as key contributors to the star-formation boom during this epoch \citep[e.g.][]{BargerEtal1998, DaddiEtal2004, MagnelliEtal2013, CaseyEtal2014}. Such systems, commonly identified as high-$z$ ultraluminous infrared galaxies (ULIRGs, with $L_{IR} > 10^{12}L_\sun$), typically have stellar masses greater than $10^{11} M_\sun$ and are considered progenitors of the most massive elliptical galaxies in the present day. Unlike their local counterparts, which are predominantly major mergers, high-$z$ ULIRGs display a diversity of formation tracks, including major mergers, minor mergers, or just simply non-merging disk galaxies \citep[e.g.][]{KartaltepeEtal2012, LingYan2022, ChengEtal2022, ChengEtal2023}.

A primary avenue for investigating high-$z$ ULIRGs involves observations in the far-infrared to submillimeter bands, spanning wavelengths from $200\micron$ to $1mm$. These submillimeter galaxies (SMGs) exhibit strong rest-frame $\sim 100\micron$ emission from large quantities of cold dust heated by newly formed stars \citep{SN1991, BargerEtal1998, PopeEtal2008} and display pronounced negative K-corrections \citep{BL1993}. Early attempts to measure their clustering using {\em SCUBA} $850\micron$ surveys yielded tentative detections based on several dozen SMGs \citep{ScottEtal2002, WebbEtal2003, BorysEtal2003, BlainEtal2004, ScottEtal2006}. Significant clustering signals were subsequently measured in the $870\micron$ LABOCA ECDFS Submillimeter Survey \citep{WeisEtal2009}, and later confirmed by multiple works analyzing samples comprising $10^2-10^3$ objects drawn from different submillimeter surveys \citep[e.g.,][]{MaddoxEtal2010, CoorayEtal2010, WilliamsEtal2011}. Clustering analysis indicates that SMGs are very likely to reside in halos with masses of $10^{12}-10^{13}M_\sun$ at $z \sim 2$ \citep[e.g.][]{HickoxEtal2012, ChenEtal2016, WilkinsonEtal2017, AmvrosiadisEtal2019, StachEtal2021}.

An equally primary method for selecting ULIRGs at $z \sim 2$ relies on near- and mid-infrared observations, particularly using the $24\micron$ data from the {\em Spitzer Space Telescope} \citep{WernerEtal2004}, although samples produced in this way would contain a diverse galaxy population and require complex photometric constraints. Prodigious star formation and AGN activity are the two major energy sources powering the infrared emission of gas-rich systems shrouded in dust \citep[e.g.][]{GenzelEtal1998, PapovichEtal2006, WeedmanEtal2006}. It is already known that the population of Dust Obscured Galaxies (DOGs) selected by $24\micron$ flux and $R-[24]$ colors actually includes a considerable, if not majority, portion of AGNs \citep{FioreEtal2008, DeyEtal2008}. As well, $z\sim 2$ star-forming BzKs \citep{DaddiEtal2004, DaddiEtal2005} suffer from the same problem \citep[e.g.][]{DaddiEtal2005, HuangEtal2009,AlexanderEtal2011, RangelEtal2013, HuangEtal2021, Liangetal2024}. The confusion by AGNs thus plagues some clustering analysis of high-$z$ ULIRGs \citep[e.g.][]{MagliocchettiEtal2008, StarikovaEtal2012}.

In principle, selection criteria together with an infrared flux cut should ensure a clear identification of star-formation-dominated high-$z$ ULIRGs, as well as be able to discriminate those mainly driven by AGNs. One effective strategy for selecting high-$z$ ULIRGs is to combine the $24\micron$ flux cut with near-infrared colors. The strong emission features of polycyclic aromatic hydrocarbon (PAH) serve as reliable tracers of intense star formation activity \citep[e.g.][]{GenzelEtal1998, RigopoulouEtal2006, SajinaEtal2007, PopeEtal2008, ShipleyEtal2016, Li2020}. For $z\sim 2$ ULIRGs, their $\sim 8\micron$ PAH features in the rest frame are redshifted to the observed $24\micron$ band. Additionally, the $1.6\micron$ bump attributed to $H^-$ ions in the atmospheres of giant stars in galaxies is searched to constrain redshift photometrically \citep{SimpsonEisenhardt1999, Sawicki2002}.
By leveraging the four {\em Spitzer}/IRAC bands ($3.6-8\micron$) data, ULIRGs can be picked up within a narrow range of $z\sim 2$ and reduced AGN contamination \citep[e.g.][]{HuangEtal2004, WeedmanEtal2006, FarrahEtal2008, HuangEtal2009, FangEtal2014}. For example, \citet{FarrahEtal2006} constructed two ULIRGs samples in this way, by first applying a $F_{24\micron} > 400\mu Jy$ selection from the {\em Spitzer} Wide-area Infrared Extragalactic Survey \citep[SWIRE;][]{LonsdaleEtal2003}, and then using IRAC colors to isolate 1689 objects in $1.5<z<2.0$ with $F_{3.6} < F_{4.5}>F_{5.8}>F_8$ and 1223 sources in $2.2<z<2.8$ with $F_{3.6}< F_{4.5}<F_{5.8}>F_8$. Angular two-point correlation functions (ACFs) of these two samples infer that ULIRGs occupy halos with masses of $10^{13.5}-10^{13.9}M_\sun$ at lower redshift and $10^{13.7}-10^{14.1}M_\sun$ at higher redshift, respectively. 

\begin{figure*}
\resizebox{\hsize}{!}{\includegraphics{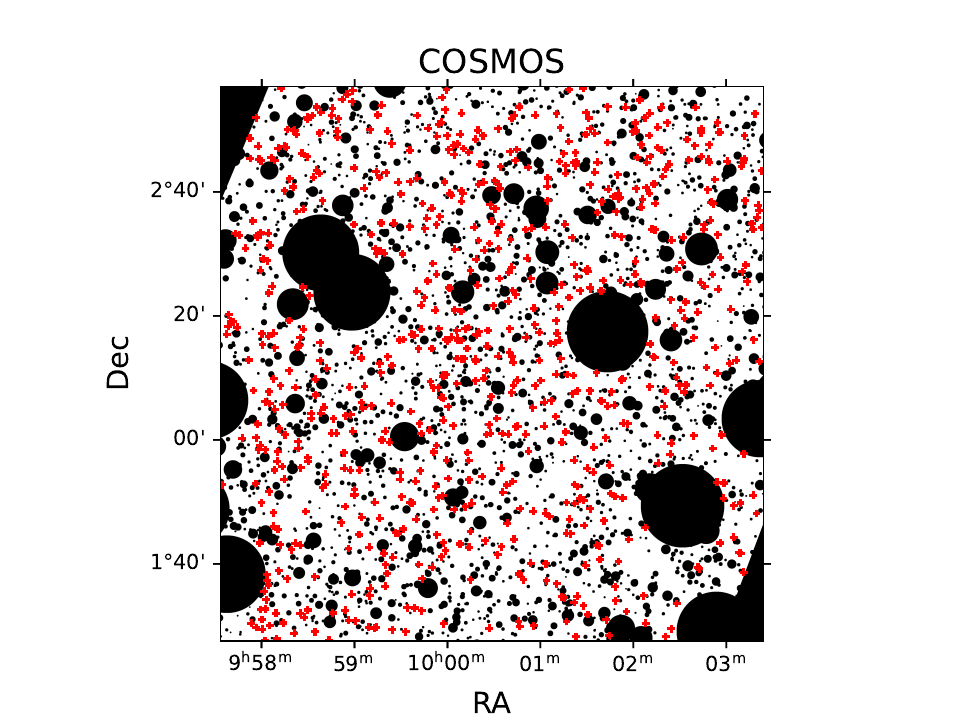}\includegraphics{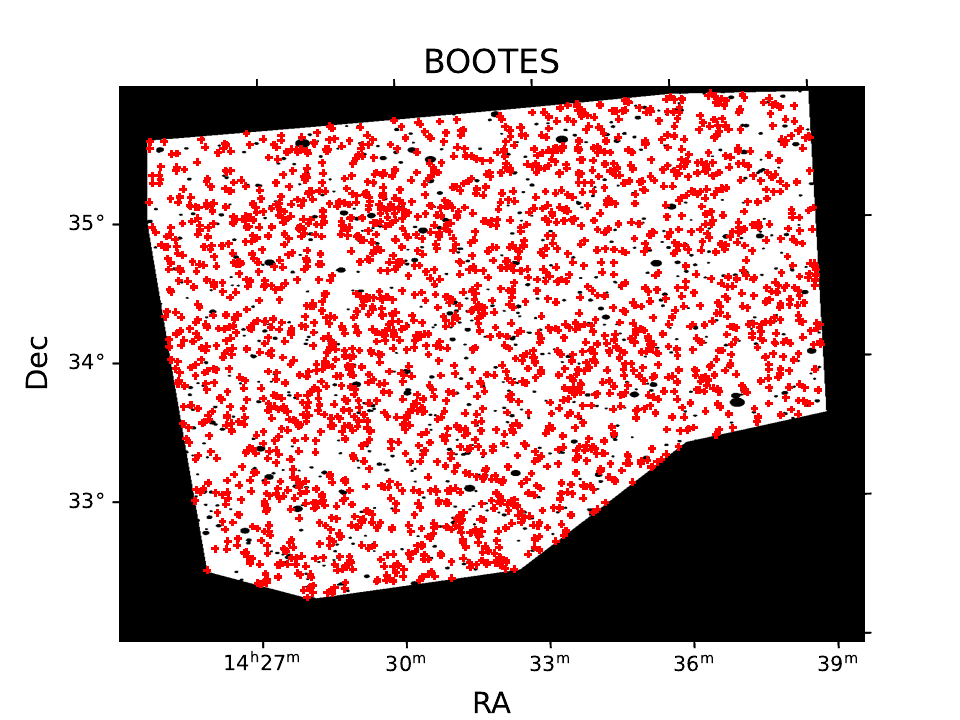}}
\resizebox{\hsize}{!}{\includegraphics[width=0.8\textwidth]{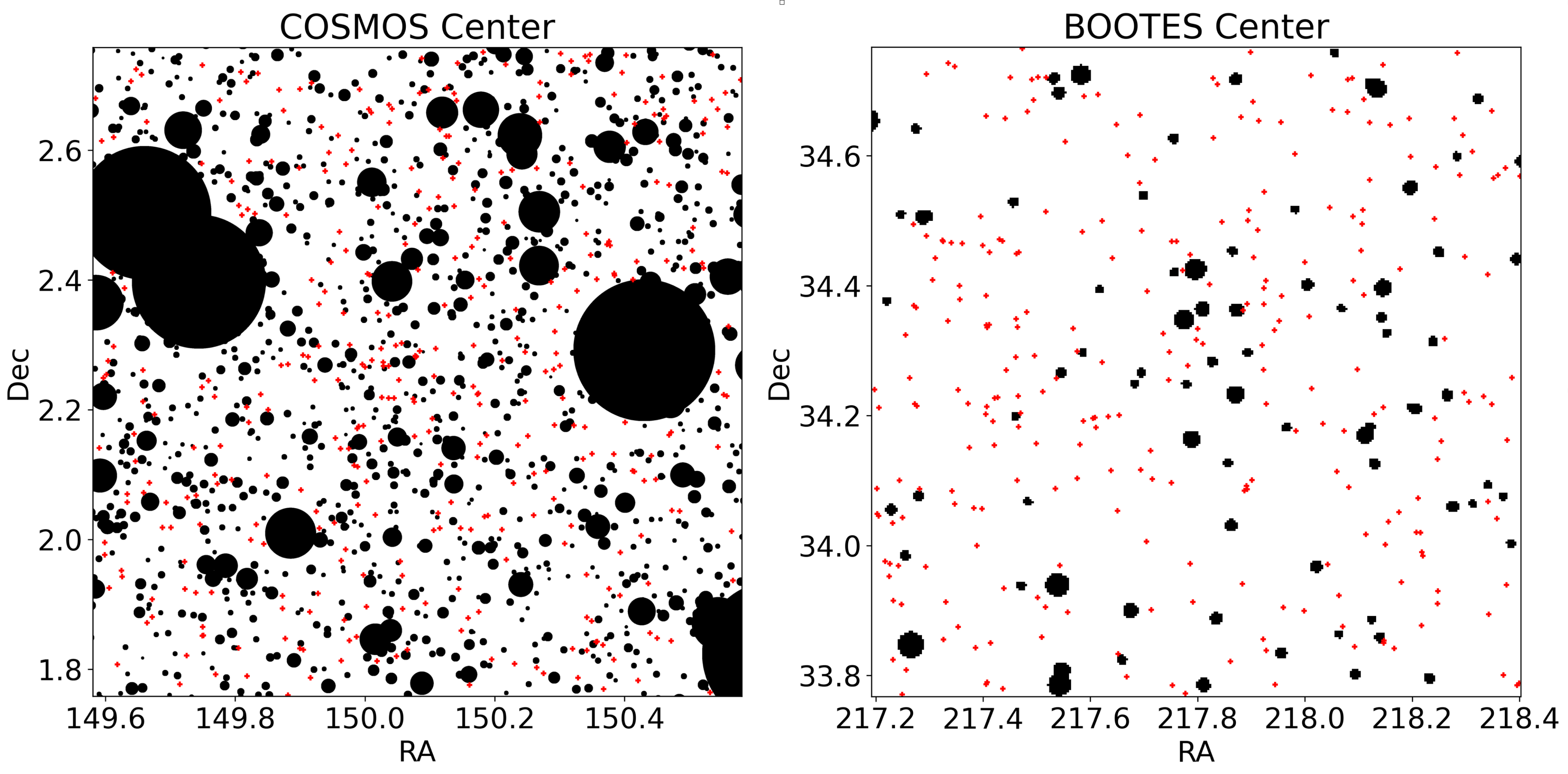}}
\caption{Sample galaxies (red symbols) overlaid with masks (black filled regions). The upper left panel shows 722 galaxies in a $1.585 \ \text{deg}^2$ region of the COSMOS field; the upper right panel shows 2268 objects distributed in the BOOTES field spanning over $8.415\ \text{deg}^2$. The bottom panels show the corresponding cutout of the central $2\ \text{deg}^2$ regions of the two fields.}
\label{fig:galpos}
\end{figure*}

In this work, we present a clustering analysis of a $z\sim 2$ ULIRG sample produced with the implementation of \citet{FangEtal2014}, with a slightly further refinement of the techniques developed by \citet{HuangEtal2004} and \citet{HuangEtal2009} to better mitigate AGN confusion. This is the largest ULIRG sample at $z\sim2$ ever used for this kind of study. Instead of using the correlation length as the primary indicator of clustering strength, as adopted in several prior works, we opt to directly fit the ACFs with models of linear biasing and the halo occupation distribution (HOD), aiming to robustly constrain the typical host halo mass of ULIRGs at cosmic noon. The sample construction and ACF estimation are described in Section 2. The analysis based on the linear bias model is presented in Section 3, while HOD fittings are discussed in Section 4. A discussion is provided in the last section. 

Throughout this work, we assume a $\Lambda {\rm CDM}$ cosmological model of $\Omega_m=0.316$, $\Omega_\Lambda=0.684$, $h=0.673$ and $\sigma_8=0.812$ compatible with the best fit of \citet{PLANCK2018}.

\section{Galaxy Samples and ACFs}
\subsection{Sample Construction}
Among the sky area covered by the {\em Spitzer} MIPS $24\micron$ and IRAC observations, two far-separated fields are chosen as the basis for our samples. One is in the COSMOS field, covering approximately $1.585 \ {\rm deg}^2$ \citep{2007ApJS..172...86S}. The corresponding  MIPS 24 $\mu$m catalog is available from the COSMOS dataset archive \footnote{\url{https://irsa.ipac.caltech.edu/data/COSMOS/gator_docs/scosmos_mips_24_go3_colDescriptions.html}}. The other is the BOOTES field, with an area of $8.415\ {\rm deg}^2$ \citep{2004AAS...204.4805S}. The catalog can be obtained from the Spitzer Data Fusion database \footnote{\url{https://www.mattiavaccari.net/df/}} \citep{2015fers.confE..27V}. The effective sky areas covered by the two survey footprints at $z\sim2$ are $\sim 3.78\ {\rm Mpc^2}$ and $20.1\ {\rm Mpc^2}$ accordingly.

We follow \citet{FangEtal2014} in selecting reliable ULIRGs within a narrow range at $z\sim 2$. Sources with MIPS flux $F_{24 \micron} > 350\rm \mu Jy$ enter our sample if they comply with the following IRAC color criteria,
\begin{equation}
[3.6]-[4.5]>0\ \ \& \ \ [5.8]-[8.0] <0\ .
\end{equation}

The $F_{24\micron}$ flux limit is for the selection of ULIRGs with strong $7.7\micron$ PAH emission at $z\sim2$. The IRAC color criteria are set to detect the $1.6\micron$ stellar bump shifting in the $4.5-5.8\micron$ bands and also to screen out power-law AGNs. Mid-infrared spectroscopy by \citet{HuangEtal2009, FangEtal2014} confirmed that this selection method was very effective in detecting PAH emissions at $z\sim2$.

The final selection yields $722$ objects in the COSMOS field and $2268$ galaxies in the BOOTES field (Fig.~\ref{fig:galpos}). The minimum angular separation between selected galaxies is $\sim 5''$. We noticed that the surface densities of objects in the two fields differ, with values of 457 deg$^{-2}$ and 270 deg$^{-2}$, respectively. This may be attributed to cosmic variance. Actually, such an effect has also been observed in the COSMOS field at lower redshifts in star-forming galaxies. For example, a potential high overdensity region at  $z\sim 0.7$ could locally boost the galaxy number \citep[]{dlTorreEtal2010,SaitoEtal2020}. Therefore, we performed clustering analysis in two fields separately rather than combining them together. Details of sample selection and its properties will be presented in Huang et al. (2026, to be submitted).

To accurately estimate angular correlation functions (ACFs), well-defined geometrical masks are necessary. These include regional masks that delineate the boundaries of the observed sky and bright star masks that cover areas shielded by bright stars. We adopt the Hyper Suprime-Cam (HSC) survey mask for the COSMOS field, where the mask radius for each bright star is determined from its G-band flux \citep[see details in][]{2018PASJ...70S...7C}. The same masking method is also applied to the BOOTES field. The results are shown in Fig.~\ref{fig:galpos}.

An essential component in modeling ACFs is the radial distribution of galaxies. Traditionally, this redshift distribution is derived from a representative subsample with spectroscopic redshifts (spec-$z$). However, substantial dust extinction in the $24\micron$ bright sample results in faint optical magnitudes, leading to incomplete spec-$z$ coverage from the optical spectrograph. To address this issue, 
we compile a small sample of $\sim 100$ ULIRGs with spectroscopic redshifts in the EGS field (with a field size of 0.48 deg$^2$), selected according to the same selection criteria as the primary sample \citep{HuangEtal2009, FangEtal2014}. The resulting spec-$z$ distribution is then binned for our analysis. We assume that the spec-$z$ distribution of the EGS $24\micron$ sample is representative of those in the COSMOS and BOOTES fields. 

\begin{figure*}
\includegraphics[height=0.35\textwidth]{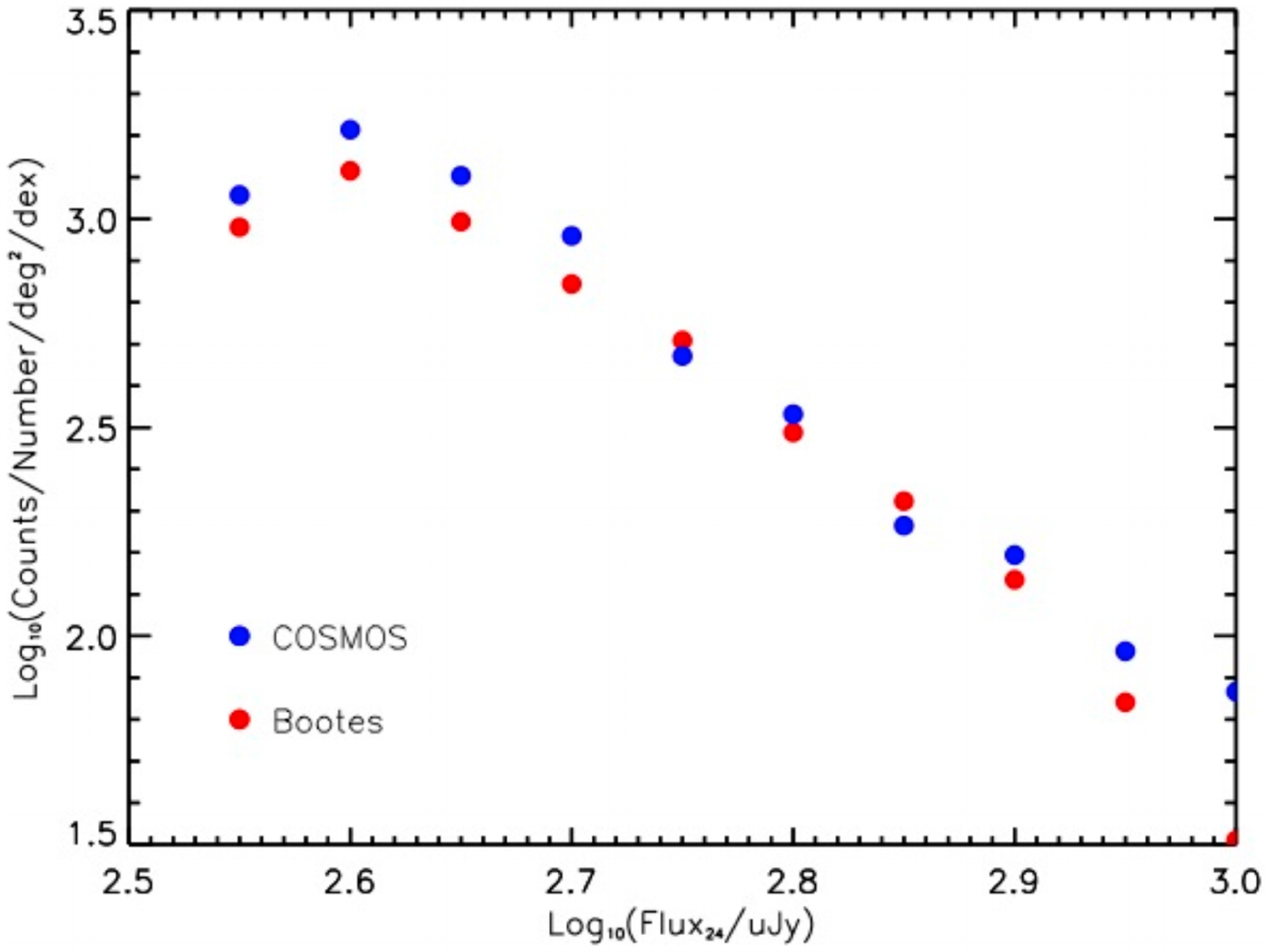}
\includegraphics[height=0.35\textwidth]{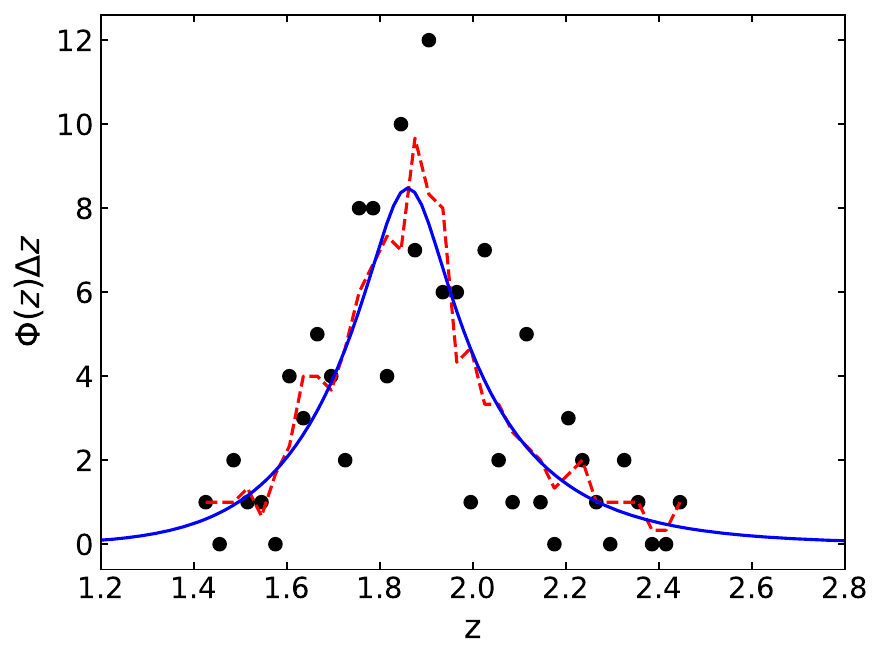}
\caption{Left panel: The observed $24\,\micron$ apparent-flux distributions of the two samples, normalized by the total number of sources and the field coverage. Right panel: Radial distribution of the ULIRG sample used to model the radial selection function. Black points are the counts of galaxies in redshift bins of width $\Delta z = 0.03$. The dashed line shows the distribution smoothed over three consecutive $z$ bins, while the solid line shows the model of Eq.~\ref{eq:nz} with best-fit parameters $(z^*, p_0,p_1,p_2,p_3)=(1.863\pm 0.013, 353.2\pm38.2, -9.74\pm1.13, 70.6\pm45.6, 22.7\pm1.7)$ and a reduced chi-square of 0.39.}
\label{fig:nz}
\end{figure*}

Since our comparison between COSMOS and BOOTES assumes a common redshift distribution, we compare the observed $24\,\micron$ apparent-flux distributions. After normalizing by the total number of sources and the field coverage (left panel of Fig.~\ref{fig:nz}), the two distributions are broadly similar. This comparison, by itself, does not prove identical redshift distributions, but it shows that the two samples do not exhibit a strong mismatch in their observed $24\,\micron$ flux distributions. This supports the use of a common redshift selection function as a practical treatment.

The binned radial distribution exhibits significant noise due to the limited sample size. To suppress local fluctuations, we parametrize the radial distribution in the right panel of Fig.~\ref{fig:nz}, using a double power law \citep{SaundersEtal2000}
\begin{equation}
\Phi(z)\rmd z = p_0 \left( \frac{z}{z^*} \right)^{1-p_1} \left[ 1+\left(\frac{z}{z^*}\right)^{p_2} \right]^{-p_3/p_2} \rmd z\ .
\label{eq:nz}
\end{equation}
The model parameters are estimated via standard $\chi^2$ minimization, assuming Poisson uncertainties in the binned data. The resulting fitted value of $z^*=1.86$ defines the characteristic redshift of our ULIRG sample. The radial selection function as a function of the comoving distance $r$ is then derived by $\phi(r) r^2 \rmd r \propto \Phi(z) \rmd z$. We note that the selection function is used to compute the theoretical models presented in Sections~\ref{sec:linmodel} and \ref{sec:hodmodel}; in doing so, we neglect the evolution of galaxy properties over the redshift range covered by our sample. As shown by \citet[]{Zheng2004}, the theoretical modeling is only weakly sensitive to the exact form of the selection function. We therefore adopt this practical parametrization rather than introducing additional degrees of freedom that cannot be constrained by the current data.

\subsection{Estimation of ACF and Covariance Matrix}
The Landy--Szalay (LS) estimator \citep{LandySzalay1993} is adopted to measure the raw angular correlation function (ACF) $\hat{\omega}(\theta)$,  
\begin{equation}
\label{eq:a2pcf_ls}
\hat{\omega}(\theta)= \frac{DD-2DR+RR}{RR}\ ,
\end{equation}
using auxiliary random catalogs containing at least $N_{\rm r} \gtrsim 10^5$ random points. The number is usually chosen to be a factor of $\sim 100$ of the data points $N_{\rm g}$, \citep[see, e.g.][]{NorbergEtal2009}, and in order to further suppress variance induced by finite $N_{\rm r}$, 200 random catalogs are employed for each run of measurement and results are combined by the method of \citet{KeihanenEtal2019} to yield the final estimation. $\theta$ denotes the angular separation, $DD$ is the number of galaxy-galaxy pairs normalized by $N_{\rm g}(N_{\rm g}-1)/2$, $DR$ is the number of galaxy-random pairs normalized by $N_{\rm g}N_{\rm r}$ and $RR$ is the number of random-random pairs normalized by $N_{\rm r}(N_{\rm r}-1)/2$. All pairs are considered distinct. The angular scales are divided into 20 bins for the COSMOS field and 30 bins for the BOOTES field, spaced logarithmically from the smallest bin at $5''$ to the maximum scale allowed by the sample geometry. 
In practice, we only use angular bins in which the raw (unnormalized) number of galaxy--galaxy pairs exceeds 10, as bins with very small pair counts are dominated by shot noise and lead to unstable estimates of $\omega(\theta)$.

For samples in a finite space, the integral constraint introduces a systematic bias in the measured ACFs. This bias can be approximately corrected by
\begin{equation}
\label{eq:ic}
\Delta \omega \equiv \omega-\hat{\omega} =\frac{\sum_i RR(\theta_i) \omega(\theta_i)}{\sum_i RR(\theta_i)}\ ,
\end{equation}
where $\theta_i$ denotes the $i^{th}$ angular bin \citep[e.g.][]{RocheEales1999}. Estimating the bias requires {\em a prior} knowledge of the true galaxy ACF, $\omega(\theta_i)$. A common approach is to assume a power-law form, $\omega=A\theta^{\gamma}$ \citep{RocheEales1999}; however, this assumption becomes inaccurate at large angular scales. A more robust method is to incorporate the integral constraint directly into the theoretical model \citep[e.g.][]{AdelbergerEtal2005}. 
In this work, we adopt the iterative procedure described by \citet{GeachEtal2012}, yielding corrections of $\Delta \omega\approx 5.036\times 10^{-3}$ for the COSMOS field and $1.634\times 10^{-3}$ for the BOOTES field (we have verified that alternative model assumptions for $\Delta\omega(\theta)$ alter these corrections by $<2\%$; see Appendix~\ref{sec:ic_test} for details).

With this correction, we fit the theoretical models to the ACF using maximum-likelihood estimation (MLE). Specifically, assuming that the ACF measurements are drawn from a multivariate Gaussian distribution and measured in $N_{\rm bin}$ angular bins, then the likelihood for obtaining the observed ACF $\hat{\vect{w}}=(\hat{\omega}(\theta_1), \hat{\omega}(\theta_2), \ldots, \hat{\omega}(\theta_{N_{\rm bin}}))$ given a set of $N_{\rm p}$ model parameters $\vect{p}=(p_1, p_2, \ldots, p_{N_p})$, is
\begin{equation}
\label{eq:likelihood}
\mathcal{L} \propto \abs{{\rm det} {\bf C}}^{-1/2}  \exp \left(- \vect{\Delta}^T \vect{\Psi} \vect{\Delta}/2\right)
\end{equation}
where the vector $\vect{\Delta}=\vect{w}_{\vect{p}}-\hat{\vect{w}}$ quantifies the difference between the model prediction and the data, and $\vect{w}_{\vect{p}}=(\omega_\vect{p}(\theta_1), \omega_\vect{p}(\theta_2), \ldots, \omega_\vect{p}(\theta_n))$ is the theoretical ACF predicted with the given parameter set $\vect{p}$. The covariance matrix is denoted by $\vect{C}$, and $\vect{\Psi}=\vect{C}^{-1}$ is the corresponding precision matrix. 
We maximize the likelihood function $\mathcal{L}$ by minimizing $\chi_{\omega}^2 = \vect{\Delta}^T \vect{\Psi} \vect{\Delta}$, neglecting the possible dependence of $\vect{C}$ on $\vect{p}$. The reduced chi-square $\chi_{\omega}^2/N_{\rm dof}$ is employed as a measure of goodness of fit, where $N_{\rm dof}=N_{\rm bin}-N_{\rm p}$.

It is well known that the correlation function in a nonlinear density field is not strictly distributed as a multivariate Gaussian, and adopting a Gaussian likelihood can sometimes underestimate uncertainties in high-precision parameter inference \citep[e.g.][]{HahnEtal2019}. However, given the sample size and signal-to-noise ratio of the current ULIRG data, the Gaussian likelihood provides an approximate but practical framework that is adequate for comparing the two fields at the precision level of this work.

The covariance matrix required for the MLE analysis is estimated using the jackknife resampling technique. For each field, we divided the galaxy sample into $N_{\rm jk}$ non-overlapping regions of approximately equal area. Each jackknife realization is constructed by systematically omitting one region at a time from the full sample. The number of jackknife regions must be large enough to provide a stable covariance estimate. We therefore tested $N_{\rm jk}=36,49,64,81,100$, and 121. We adopt $N_{\rm jk}=100$ for COSMOS and $N_{\rm jk}=81$ for BOOTES as our fiducial choices (for more technical details, see Appendix~\ref{sec:jk_num}). Then the ACFs computed from these subsamples are used to estimate the covariance matrix:
\begin{equation}
\hat{C}_{ij} = \frac{N_{\rm jk}-1}{N_{\rm jk}} \sum_{k=1}^{N_{\rm jk}} (\hat{\omega}_{i, k}-\avg{\hat{\omega}_i})(\hat{\omega}_{j,k} - \avg{\hat{\omega}_j})
\label{eq:covmat}
\end{equation}
with $\avg{\hat{\omega}_i} = \sum_{k=1}^{N_{\rm jk}} \hat{\omega}_{i,k}/N_{\rm jk}$, and $\omega_{i,k}$ is the ACF value in bin $\theta_i$ for the $k^{th}$ jackknife subsample. To correct for the bias introduced by the finite number of jackknife samples $N_{\rm jk}$, we apply the factor $(N_{\rm jk}-N_{\rm bin}-2)/(N_{\rm jk}-1)$ to  $\hat{\bf C}^{-1}$ following the prescription of \citet{HartlapEtal2007}, to obtain an unbiased estimate of the precision matrix $\vect{\Psi}$.

An alternative to jackknife resampling is to compute an analytic covariance matrix under a Gaussian random field assumption (the so-called Gauss--Poisson model; e.g.\ \citealt{GriebEtal2016}). On the quasi-linear scales where our linear model is fitted ($\theta \gtrsim 1'$), the nonlinear contribution to the ACF is less than $\sim10\%$; such an analytic approach would be expected to be accurate. However, implementing the Gauss--Poisson covariance for an ACF requires accurate knowledge of the survey window function. The jackknife method automatically accounts for the survey geometry and irregular masks, and has been shown to produce covariance estimates consistent with those from mock catalogs \citep[e.g.,][]{NorbergEtal2009}. We therefore regard the jackknife covariance as the fiducial choice for the present data set. Since the present analysis is limited by sample size and survey geometry rather than by percent-level covariance modeling, we do not expect this choice to qualitatively affect the inferred difference between the two fields. A full comparison with mock-based or analytic covariance matrices will be valuable for future larger ULIRG samples.

\begin{figure*}
\plottwo{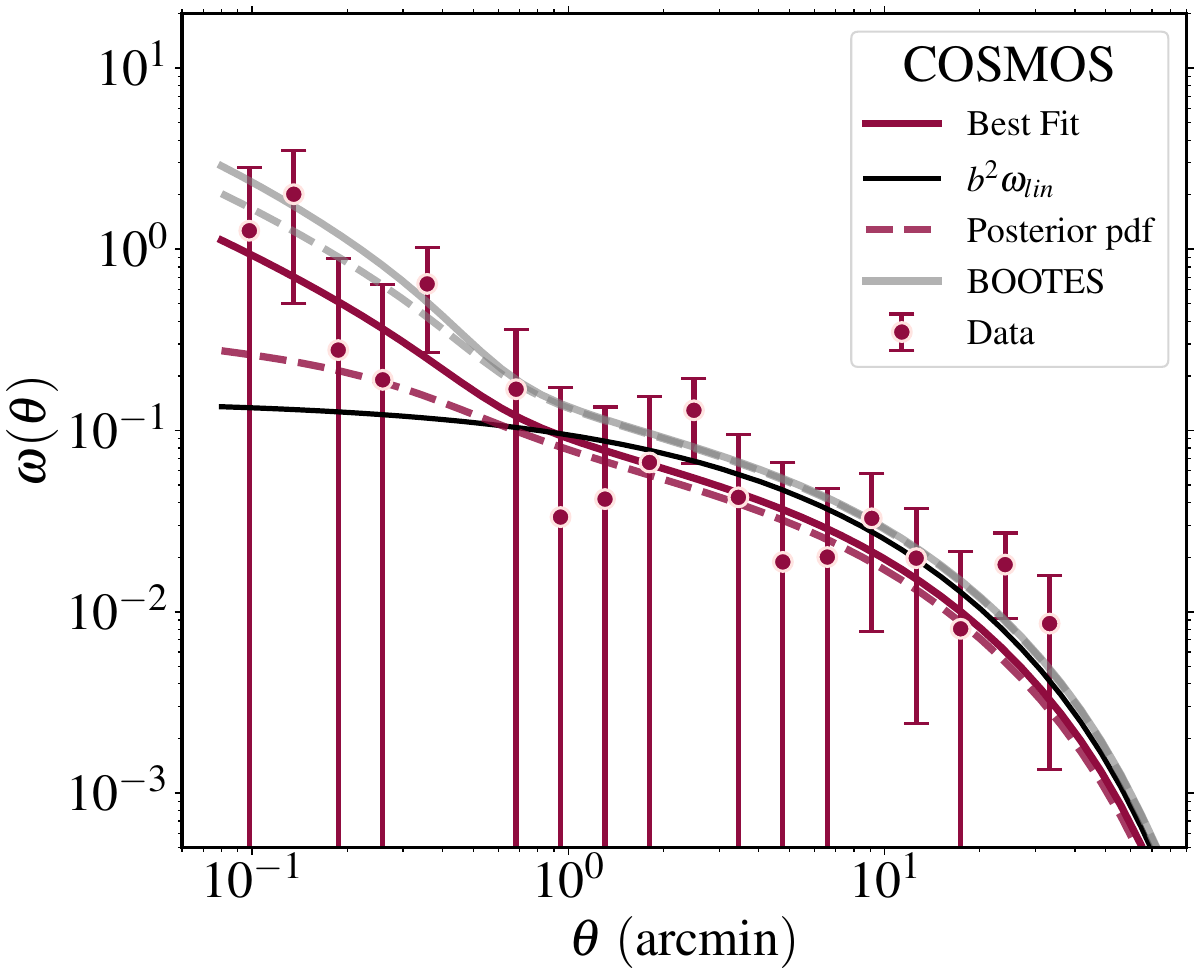}{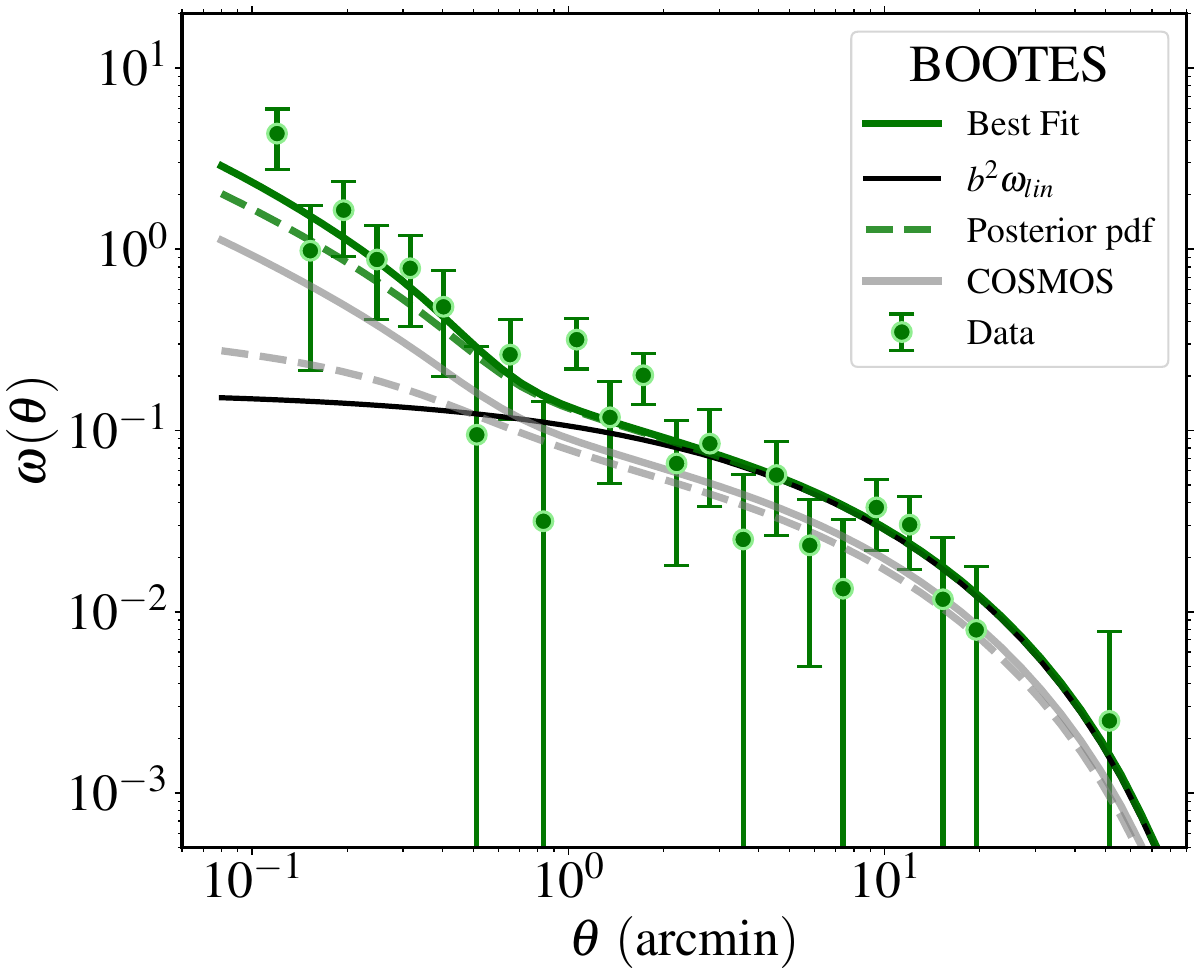}
\caption{Angular correlation functions (ACFs) of ULIRGs at $z\sim 2$ in the COSMOS (left panel) and BOOTES (right panel) fields, respectively (filled circles with 1 $\sigma$ error bars). The solid color lines represent the best-fit HOD model, while the dashed color lines denote the HOD model evaluated at the posterior-median parameter values. Black solid lines are the linear model with best-fit linear bias $b_{\rm g}$ parameter presented in Table~\ref{tab:linear}. To facilitate comparison between the two fields, the fitted HOD model from the other field is overplotted as grey lines.}
\label{fig:acf}
\end{figure*}

With the covariance matrix, integral constraint, and likelihood framework established above, we now proceed to model the measured ACFs. Figure~\ref{fig:acf} presents the COSMOS and BOOTES angular correlation functions with $1\sigma$ jackknife uncertainties. The colored solid curves show the best-fitting models described below; the dashed curves in each panel show the HOD prediction evaluated at the posterior-median parameters, and the grey curves show the fitted model obtained from the other field for direct comparison.

\section{Linear model}
\label{sec:linmodel}
\begin{deluxetable*}{l c c c c c c c c}
\tablecaption{Samples and  parameters of the linear model}
\label{tab:linear}
\tablehead{ \colhead{Field} &  \colhead{area (${\rm deg}^2$)} & $N_{\rm g}$  & \colhead{($\theta_{\rm min}, \theta_{\rm max}$)} & \colhead{$N_{\rm jk}$} & \colhead{$\Delta \omega$} & \colhead{$b_{\rm g}$}& \colhead{$\log M_{\rm h}/(\msun)$} }
\startdata
(1) & (2) & (3) & (4) & (5) & (6) & (7) & (8) \\
COSMOS & $1.585$ & $722$   & $(1.31^\prime, 24.04^\prime)$ & $100$ & $5.04\times 10^{-3}$ & $3.04 \pm 0.54$ & $12.64 \pm 0.50$\\
BOOTES & $8.415$ & $2268$  & $(1.35^\prime, 51.37^\prime)$ & $ 81$ & $1.63\times 10^{-3}$ & $3.22 \pm 0.28$ & $12.72 \pm 0.13$
\enddata
  \begin{tablenotes}
    \item (1): field name; (2)-(3): sky area and ULIRG number; 
    \item (4): angular scale bound when fitting to the model;
    \item (5)-(6): the number of jackknife samples and the integral constraint correction;
    \item (7)-(8): the fitting-derived bias parameter and inferred halo mass for ULIRGs.
  \end{tablenotes}
\end{deluxetable*}

In this section, we first derive the theoretical ACF within the framework of the linear model. We begin by describing the spatial distribution of galaxies in terms of fluctuations about the mean density field. The number density of galaxies at position $\vect{r}(z)$ can be written as $n_{\rm g}(\vect{r}) = \bar{n}_{\rm g}(r)\,[1+\delta_{\rm g}(\vect{r})]$, where $\bar{n}_{\rm g}(r)$ is the mean number density at redshift $z$, and $\delta_{\rm g}(\vect{r})$ denotes the local density contrast relative to $\bar{n}_{\rm g}(r)$. The three-dimensional two-point correlation function (2PCF) is then defined as $\xi_{\rm g}(\vect{r}_1,\vect{r}_2)=\langle \delta_{\rm g}(\vect{r}_1)\delta_{\rm g}(\vect{r}_2)\rangle$. At large scales, a linear model is expected to provide an adequate description of galaxy clustering. The {\em linear} here means that (i) the 2PCF of dark matter can be well approximated by the linear theory, and (ii) the galaxy bias is a constant and scale-independent. Under these assumptions, the 2PCF of galaxies can be expressed as $\xi_{\rm g}=b_{\rm g}^2 (D_{z}/D_0)^2 \xi_0$, where $b_{\rm g}$ is the linear bias, $D_z$ is the linear growth factor at $z$, and $\xi_0$ represents the linear 2PCF of dark matter extrapolated to $z=0$.

If correlations between galaxy properties and their spatial positions are negligible, and clustering evolution of galaxies across the sample's redshift range is insignificant \footnote{We performed a rigorous calculation showing that linear evolution with redshift contributes less than $1.3\%$ variation to $\omega$.}, we can describe the ACF using the linear model with Limber's equation \citep{Limber1953, Peebles1980},
\begin{equation}
\omega(\theta)  = \frac{\int_0^{+\infty}  \phi^2(x) x^4 \int_{-\infty}^{+\infty}  b_{\rm g}^2 \frac{D^2_{z^*}}{D^2_0}\xi_0(\sqrt{y^2+x^2\theta^2})  \rmd y  \rmd x}{ \left[   \int_0^{+\infty} \phi(x) x^2  \rmd x \right]^2} \ ,
\label{eq:linearmodel}
\end{equation}
where $\phi(x) x^2 \rmd x \propto \Phi(z)\rmd z$ and $z^*=1.86$ is the characteristic redshift.

The simple way to infer the mass of the host halo $M_{\rm h}$ is to assume that the linear bias of galaxies is solely determined by their host halos, so that $b_{\rm g}=b_{\rm h}(M_{\rm h})$. The halo mass obtained in this way thus represents an {\em effective} mass for the galaxy sample. For this purpose, we adopt the bias-mass relation from \citet{TinkerEtal2010},
\begin{equation}
b_{\rm h}(\nu)= 1-A \frac{\nu^a}{\nu^a +\delta_c^a} +B \nu^b +C \nu^c\ ,
\label{eq:blin}
\end{equation}
the parameters $(A, a, B, b, C, c)$ depend on the overdensity threshold $\Delta$ \citep[see Table 2 of][here we choose $\Delta =200$.]{TinkerEtal2010}, the density peak height $\nu = \delta_c/\sigma(M_{\rm h})$, is defined with the critical density of spherical collapse $\delta_c$ and the linear matter variance $\sigma(M_{\rm h})$ on the Lagrangian scale of the halo $R=[3M_{\rm h}/(4\pi \rho_m)]^{1/3}$ at $z^*$. 

Taking into account the applicable range of the linear model and the size of our data set, we put angular scale bounds $(\theta_{\rm min}, \theta_{\rm max})$ to maximize the quality of fitting of the model. We set the upper angular limit, $\theta_{\rm max}$, to no more than $\sim 1/4$ of the sample size, since including data beyond this scale might seriously spoil the analysis. Setting the lower angular cutoff, $\theta_{\rm min}$, is more crucial. It labels the scale above which the linear model remains accurate. To determine $\theta_{\rm min}$, we calculated the nonlinear ACF of dark matter by replacing $D^2(z^*)/D^2(0)\xi_0$ in Eq.~\ref{eq:linearmodel} with the nonlinear 2PCF $\xi_{\rm nl}(z^*)$ computed with the {\tt HMcode} \citep{MeadEtal2021}. Then $\theta_{\rm min}$ is chosen to be around $1.3192^\prime$ at which the deviation between the nonlinear and the linear ACFs remains below $10\%$. 

The results of our MLE fitting to the linear model are presented in Fig.~\ref{fig:acf} (black solid lines) and summarized in Table~\ref{tab:linear}. The two fields have markedly different galaxy surface densities. Although the BOOTES field has a lower surface density than COSMOS (270 deg$^{-2}$ versus 457 deg$^{-2}$), its ACF shows somewhat smaller uncertainties, likely because the larger BOOTES area reduces sample variance on the angular scales considered. Nevertheless, the ACFs at large angular scales, as well as the corresponding derived bias parameters, are in good agreement within quoted uncertainties. The typical halo mass for ULIRGs in our samples is $\sim10^{12.7} \msun$ (i.e., $\sim 5\times 10^{12} \msun$), approximately one order of magnitude lower than the previous result reported by \citet{FarrahEtal2006}. 

\section{Halo Occupation Distribution} 
\label{sec:hodmodel}

\begin{table*}
\begin{center}
\begin{threeparttable}
\caption{Parameters and inferred quantities of the HOD model}
  \begin{tabular*}{\textwidth}{ll rr rr}
    \toprule
    & & \multicolumn{2}{c}{{COSMOS}} & \multicolumn{2}{c}{{BOOTES}}   \\ 
    \cmidrule(rl){3-4}  \cmidrule(rl){5-6}
    Parameters  & \multicolumn{1}{c}{Prior \tnote{a}} & {\tt Posterior PDF} \tnote{b} & \multicolumn{1}{c}{{\tt Best fit} \tnote{c}} & {\tt Posterior PDF} & \multicolumn{1}{c}{{\tt Best fit}}\\
    \midrule
    $\log{M_{\rm min}} / (\msun)$  & \multicolumn{1}{l |}{$[12.5, 16.5]$}  & $14.15_{-0.72}^{+0.99}$  &  \multicolumn{1}{r}{$13.53$}  & $13.52_{-0.31}^{+0.37}$  &  \multicolumn{1}{r}{$13.57$}  \\ 
    $\log{M_1}         / (\msun)$  & \multicolumn{1}{l |}{$[12, 20]$}      & $16.22_{-2.00}^{+2.56}$  &  \multicolumn{1}{r}{$15.01$}  & $14.99_{-0.50}^{+1.27}$  &  \multicolumn{1}{r}{$14.91$}  \\
    $\log{M_0}         / (\msun)$  & \multicolumn{1}{l |}{$[4, 20]$}       & $12.09_{-5.25}^{+5.53}$  &  \multicolumn{1}{r}{$ 8.36$}  & $10.48_{-3.76}^{+2.21}$  &  \multicolumn{1}{r}{$11.58$}  \\ 
    $\sigma_{\log{M}}$             & \multicolumn{1}{l |}{$[0,1.5]$}       & $ 1.25_{-0.44}^{+0.47}$  &  \multicolumn{1}{r}{$ 0.86$}  & $ 0.77_{-0.28}^{+0.23}$  &  \multicolumn{1}{r}{$ 0.82$}  \\
    $\alpha$                       & \multicolumn{1}{l |}{$1.00\pm 0.20$}  & $ 1.00_{-0.19}^{+0.19}$  &  \multicolumn{1}{r}{$ 0.98$}  & $ 1.02_{-0.20}^{+0.17}$  &  \multicolumn{1}{r}{$ 0.96$}  \\
    $\chi^2/N_{\rm dof}$ \tnote{d} & \multicolumn{1}{c |}{$\cdots$}        & \multicolumn{1}{c}{$\cdots$}&  $0.79$ & \multicolumn{1}{c}{$\cdots$}&  $0.87$ \\
    \midrule
    Inferred quantities  &   &  &  & & \\
    \midrule
    $\log n_{\rm g}^{\rm HOD} / (\lhmpc)^{-3}$ & \multicolumn{1}{c |}{$\cdots$} & $-3.97_{-0.12}^{+0.12}$ & \multicolumn{1}{r}{$-3.98$} & $-4.20_{-0.13}^{+0.13}$ & \multicolumn{1}{r}{$-4.19$}  \\
    $b_{\rm eff}$                              & \multicolumn{1}{c |}{$\cdots$} & $ 2.32_{-0.50}^{+0.65}$ & \multicolumn{1}{r}{$ 2.73$} & $ 3.25_{-0.34}^{+0.31}$ & \multicolumn{1}{r}{$ 3.29$}  \\
    $\log{M_{\rm eff}} / (\msun) $             & \multicolumn{1}{c |}{$\cdots$} & $12.50_{-0.28}^{+0.26}$ & \multicolumn{1}{r}{$12.72$} & $12.89_{-0.12}^{+0.11}$ & \multicolumn{1}{r}{$12.88$}  \\
    $f_{\rm sat} $                             & \multicolumn{1}{c |}{$\cdots$} & $ 0.00_{-0.00}^{+0.20}$ & \multicolumn{1}{r}{$ 0.26$} & $ 0.15_{-0.15}^{+0.21}$ & \multicolumn{1}{r}{$ 0.13$}  \\
    \bottomrule
  \end{tabular*}
  \label{tab:hod}
  \begin{tablenotes}
    \item[a] Pairs of numbers in brackets are intervals within which uniform priors are preset, $\alpha$ is estimated with a Gaussian prior specified by its mean and standard deviation. 
    \item[b] Parameters as medians of marginalized posterior distributions, errors are associated with $16-84$ percentiles. 
    \item[c] Parameters corresponding to the minimum of $\chi^2$.
    \item[d] Number of ACF data points used for fitting here is 18 for COSMOS and 23 for BOOTES, number of HOD parameters is 5, so $N_{\rm dof} = 14$ and $19$ for COSMOS and BOOTES, respectively.
  \end{tablenotes}
\end{threeparttable}
\end{center}
\end{table*}

As presented in Fig.~\ref{fig:acf}, the linear theory model (black solid lines) can accurately describe the ACFs only at angular scales larger than $1'$. At smaller scales $\lesssim 1'$, the nonlinear clustering signals of ULIRGs are more pronounced than predicted by a simple extrapolation of the linear dark matter ACF scaled up by $b_{\rm g}^2$. It has been generally recognized that the 2PCF of galaxies consists of two components: at large scales, clustering is dominated by correlations between galaxy pairs residing in different halos, while in the nonlinear regime, it arises mainly from objects within the same halo. To interpret the excess small-scale clustering, particularly the pronounced one-halo signal in BOOTES (Fig.~\ref{fig:acf}), we employ the halo occupation distribution (HOD) framework of \citet{ZhengEtal2005,ZhengEtal2007} (Zheng HOD model), which provides a statistical description of how galaxies populate dark matter halos. Comprehensive descriptions of this approach are available in the literature \cite[e.g.][]{BerlindEtal2002, CoorayEtal2002, ZhengEtal2005}. Here, we briefly summarize the key components relevant to our analysis. 

In the HOD formalism, the averaged number of galaxies in a halo of mass $M_{\rm h}$ at redshift $z$ can be decomposed into central and satellite contributions,  
\begin{equation}
\langle N(M_{\rm h}, z)\rangle = \langle N_{\rm c}(M_{\rm h}, z)\rangle + \langle N_{\rm s}(M_{\rm h}, z)\rangle, 
\end{equation}
where the central term is parametrized as 
\begin{equation}
\label{eq:Nc}
\langle N_\mathrm{c} (M_{\rm h})\rangle = \frac{1}{2} \left[ 1 +
  \mathrm{erf} \left( \frac{\log (M_{\rm h}/M_\mathrm{min})}{\sigma_{\log M}} \right) \right] ,
\end{equation}
and the satellite term is given by
\begin{equation}
  \langle N_\mathrm{s} (M_{\rm h})\rangle =  \langle N_\mathrm{c} (M_{\rm h})\rangle
  \left( \frac{M_{\rm h}-M_0}{M_1} \right)^\alpha .
\end{equation}
The five parameters, $\vect{p} = (\log{M_{\rm min}}, \log{M_1}, \log{M_0}, \sigma_{\log{M}}, \alpha)$ set, respectively, the central threshold, the characteristic satellite mass scale, the satellite cutoff mass, the width of the central transition, and the satellite power-law slope. 

The form of Eq.~\ref{eq:Nc} assumes that the probability for hosting a central ULIRG increases smoothly with halo mass. This is a simplified description. For low-redshift star-forming samples, the central occupation can be non-monotonic or peaked at relatively low halo masses. However, our ULIRG sample consists of very luminous infrared galaxies at $z\sim2$, where high star-formation rates and/or obscured AGN activity are preferentially associated with massive, gas-rich systems and merger-rich environments. Given the limited signal-to-noise ratio of the current ACFs, a more flexible non-monotonic HOD would introduce additional degeneracies that cannot be robustly constrained. We therefore adopt the five-parameter Zheng model as a minimal effective parametrization that captures the bulk host-halo properties while remaining sufficiently constrained by the data.

From the occupation function $\langle N (M_{\rm h}) \rangle$, we derive four physically meaningful quantities: 
\begin{equation}
\label{eq:ng}
    n_{\rm g}^{\rm HOD} = \int \rmd M_{\rm h}\ \frac{\rmd n}{\rmd M_{\rm h}}  \langle N (M_{\rm h}) \rangle \ ,
\end{equation}
    
\begin{equation}
    b_{\rm eff} = \frac{1}{n_{\rm g}^{\rm HOD}} \int \rmd M_{\rm h}\ b_{\rm h} \left(M_{\rm h}\right) \frac{\rmd n}{\rmd M_{\rm h}}  \langle N (M_{\rm h}) \rangle \ ,
\label{eq:bgal}
\end{equation}

\begin{equation}
     M_{\rm eff} = \frac{1}{n_{\rm g}^{\rm HOD}} \int \rmd M_{\rm h}\ M_{\rm h} \frac{\rmd n}{\rmd M_{\rm h}} 
        \langle N (M_{\rm h}) \rangle \ ,
\label{eq:m_hod}
\end{equation}

\begin{equation}
    f_{\rm sat} = \frac{1}{n_{\rm g}^{\rm HOD}} \int \rmd M_{\rm h} \frac{\rmd n}{\rmd M_{\rm h}}  \langle N_\mathrm{s} (M_{\rm h}) \rangle \ ,
\label{eq:fsat_hod}
\end{equation}
the model-predicted galaxy number density $n_{\rm g}^{\rm HOD}$ (Eq.~\ref{eq:ng}, with $\rmd n / \rmd M_{\rm h} $ the halo mass function), the effective large-scale bias $b_{\rm eff}$ (Eq.~\ref{eq:bgal}), the occupation-weighted effective halo mass $M_{\rm eff}$ (Eq.~\ref{eq:m_hod}), and the satellite fraction $f_{\rm sat}$ (Eq.~\ref{eq:fsat_hod}). These integrals are evaluated at the characteristic redshift $z^{*}=1.86$, and the predicted angular correlation function $\omega_{\rm p}(\theta)$ for any parameter set $\mathbf{p}$ is computed using the \texttt{HALOMOD} package \citep{MurrayEtal2021}.

We constrain the HOD parameters by minimizing $\chi^{2} = \chi^{2}_{\omega} + \chi^{2}_{\rm n}$, where the number-density term $\chi^{2}_{\rm n}$ is given by Eq.~(\ref{eq:chi2_n})
\begin{equation}
\label{eq:chi2_n}
\chi_{\rm n}^2 = [\log n_{\rm g}^{\rm obs}-\log n_{\rm g}^{\rm HOD}(\vect{p})]^2/ \sigma^2_{\log n_{\rm g}}\ .
\end{equation}
Strictly speaking, $n_{\rm g}^{\rm obs}$ and $\omega(\theta)$ are not statistically independent because both are derived from the same galaxy catalog \footnote{To test the dependence between $n_{\rm g}^{\rm obs}$ and $\omega(\theta)$, we performed a numerical experiment by computing the covariance matrix with and without the galaxy surface number density $n^{\rm s}_{\rm g}$ measured in each jackknife subsample. We compared the eigenvalues of the two matrices and found that the fractional differences are generally less than 5 percent. We therefore conclude that ignoring the covariance between $n_{\rm g}$ and $\omega(\theta)$ is a safe approximation for the present analysis.}; nevertheless, this additive $\chi^{2}$ construction is a standard practical approximation in the literature \citep{ZehaviEtal2011,KashinoEtal2017,OkumuraEtal2021}. Its inclusion is essential because it breaks a key degeneracy: without the abundance constraint, a high ACF amplitude could be explained either by a rare, highly biased population in very massive halos or by a more abundant population in moderately massive halos. The adopted uncertainty $\sigma_{\log n_{\rm g}} = 0.03\,|\log n_{\rm g}^{\rm obs}|$ follows \citet{OkumuraEtal2021}.

\begin{figure*}
\plottwo{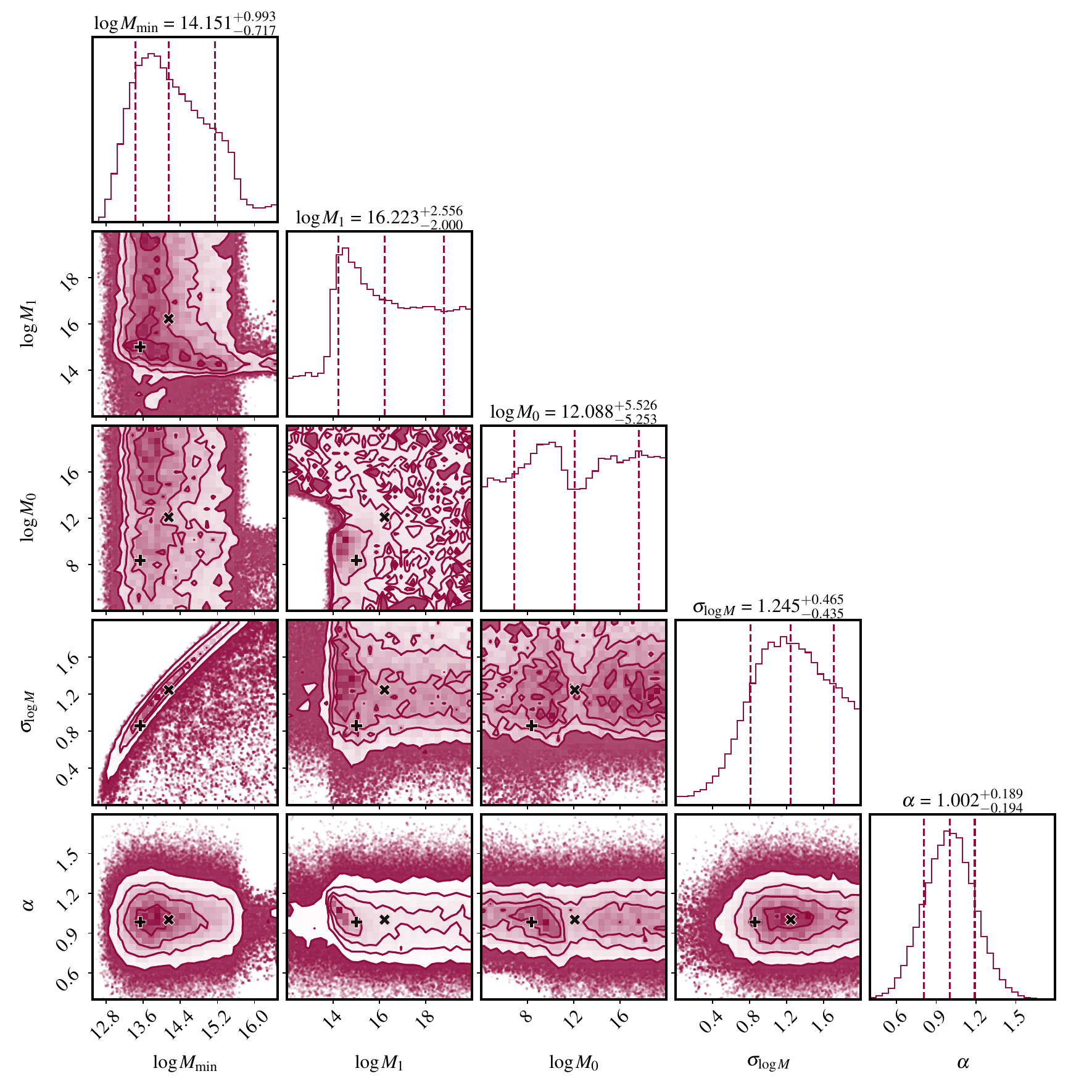}{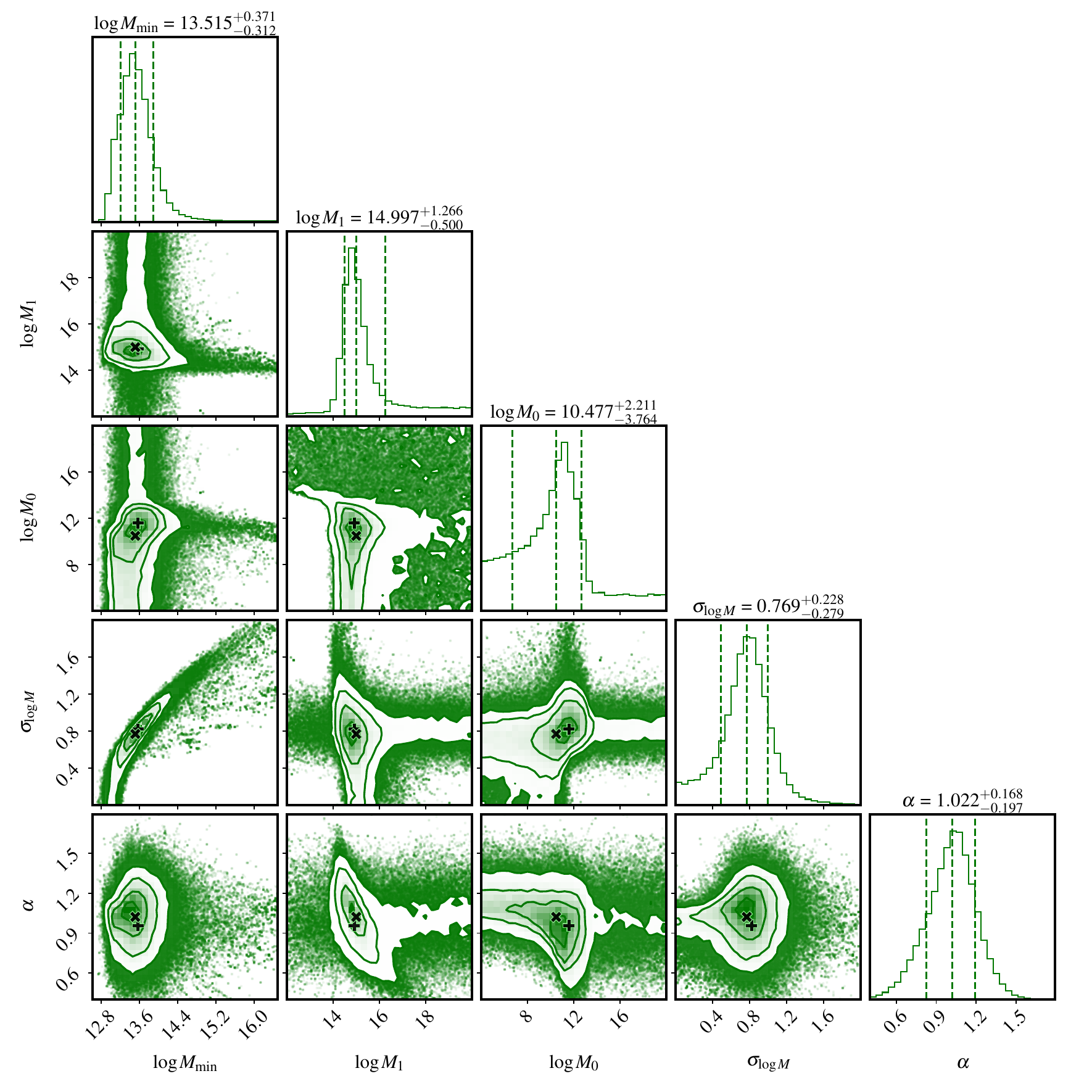}
\plottwo{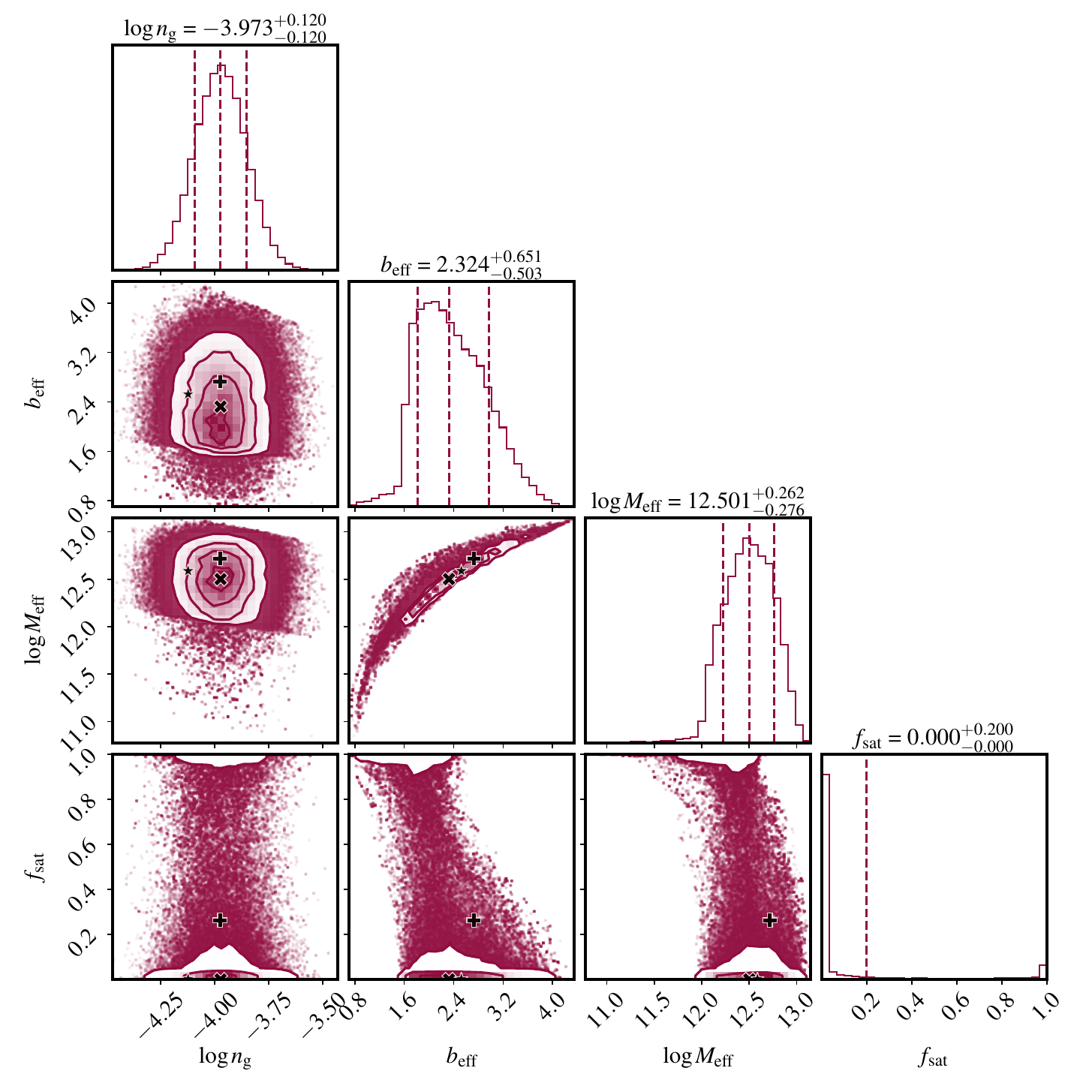}{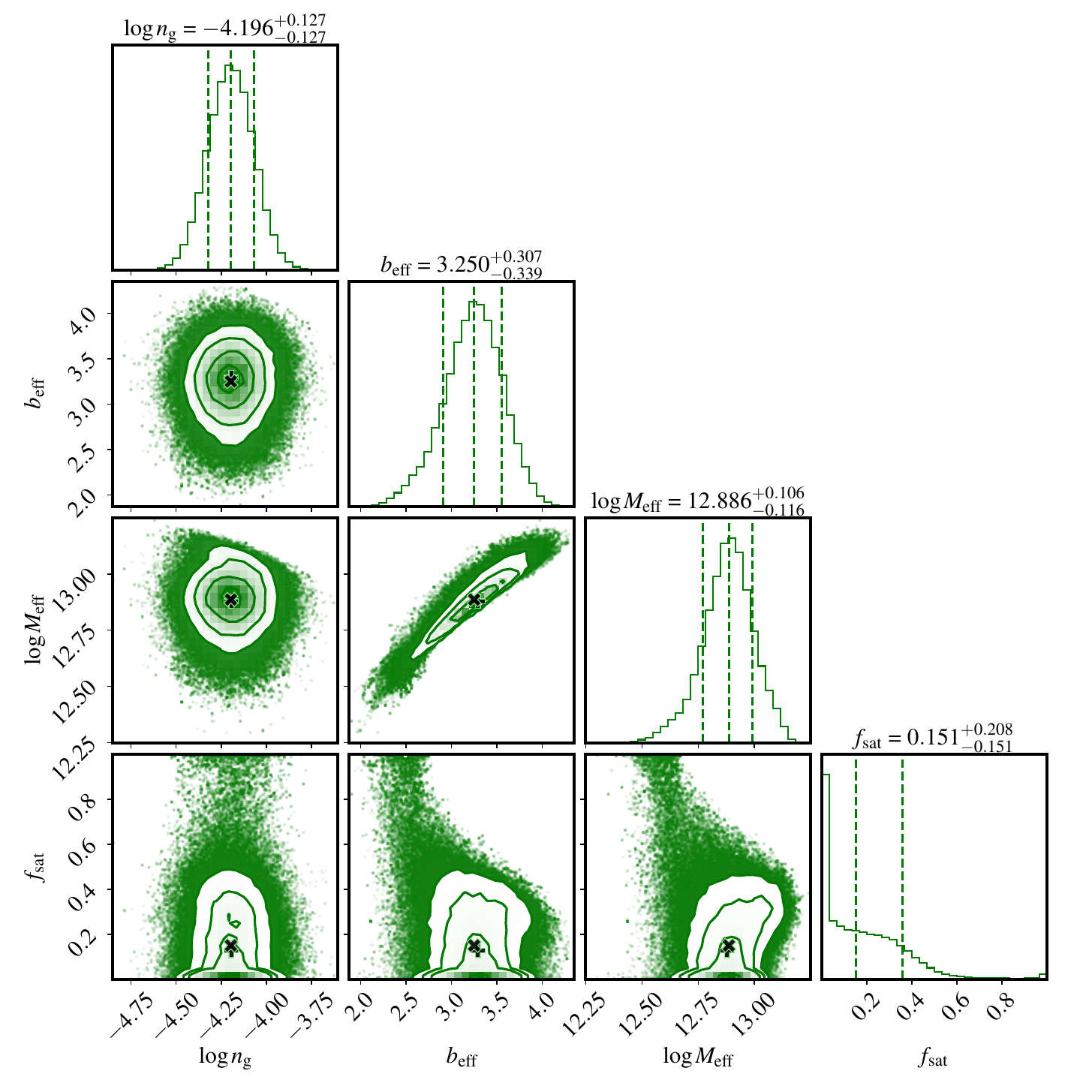}
\caption{The MCMC sampling results of Zheng's HOD model \citep{ZhengEtal2007}. Top panels: Constraints on model parameters $\vect{p} = (\log{M_{\rm min}}, \log{M_1}, \log{M_0}, \sigma_{\log{M}}, \alpha)$. Bottom panels: Posterior distributions of inferred quantities $(\log n_{\rm g}^{\rm HOD}, b_{\rm eff}, \log{M_{\rm eff}}, f_{\rm sat})$. The left column corresponds to COSMOS, and the right column to BOOTES. 
Two-dimensional contours in each panel show the 11.8\%, 39.3\%, 67.5\%, and 86.4\% confidence levels after other parameters are marginalized over. The `plus' and `cross' symbols mark the {\tt Best fit} and the {\tt Posterior PDF} values of the corresponding parameter. The diagonal sub-panels show the posterior probability distribution of each parameter, and the vertical dashed lines mark the 15.9, 50.0, and 84.2 percentiles.}
\label{fig:hod_mcmc}
\end{figure*}

The Markov chain Monte Carlo (MCMC) sampler {\tt emcee} of \citet{Foreman-MackeyEtal2013} is employed to explore the parameter space of the HOD model. Posterior distributions for both the free parameters and inferred quantities are demonstrated in Fig.~\ref{fig:hod_mcmc}. The priors and constraints on the HOD parameters are summarized in Table~\ref{tab:hod}, along with the corresponding derived values of Eqs.\ref{eq:ng} -- \ref{eq:fsat_hod}.  

Table~\ref{tab:hod} lists two sets of HOD parameter estimates, the {\tt Best fit} values corresponding to the parameter with the minimum $\chi^2$ in the MCMC chains, and the {\tt Posterior PDF} values provided by the medians of the posterior distributions after other parameters are marginalized over. For well-constrained parameters, the two agree closely; they diverge only where the data provide limited information, as also noted in previous HOD analyses \citep[e.g.,][]{HongEtal2019,OkumuraEtal2021}. Specifically, in COSMOS, the ACF is dominated by the two-halo term of central galaxies, so the one-halo signal is too weak to constrain the satellite-related parameters $M_{0}$, $M_{1}$, and $\alpha$. Conversely, in BOOTES galaxies exhibit a stronger small-scale clustering signal (see Fig.~\ref{fig:acf}) compared to COSMOS, which in turn provides somewhat more information on the satellite component of the HOD model.

The HOD-derived effective halo masses are $M_{\rm eff}=10^{12.50}\ \msun$ for COSMOS and $10^{12.89}\ \msun$ for BOOTES, with effective biases $b_{\rm eff}\sim 2.32$ and $3.25$, respectively. As expected, $M_{\rm eff}$ and $b_{\rm eff}$ exhibit a strong positive correlation in the posterior distributions of both fields (with Pearson $r\sim 0.9$), reflecting the monotonic increase of the halo bias $b_{\rm h}(M_{\rm h})$ with halo mass in Eq.~\ref{eq:blin}. This degeneracy is physically well motivated: more massive halos are more biased. This correlation should therefore be taken into account when interpreting the two quantities, rather than treating them as independent constraints.

It is notable that the marginalized $\alpha$ posteriors are similar between the two fields (Table~\ref{tab:hod}). However, this similarity should not be over-interpreted. In COSMOS, the satellite fraction is consistent with zero, so the one-halo term is essentially absent; consequently, $\alpha$ is prior-dominated, and its posterior largely reflects the Gaussian prior we imposed. In BOOTES, the stronger small-scale clustering does provide some information about the satellite occupation, yet $\alpha$ itself remains moderately constrained. Thus, the comparable $\alpha$ values do not imply physically identical satellite populations.

Combining the clustering and abundance constraints reveals a counter-intuitive but physically coherent picture. Although COSMOS has the higher surface density ($457\,{\rm deg}^{-2}$ versus $270\,{\rm deg}^{-2}$), its ULIRGs are consistent with being pure centrals ($f_{\rm sat}\sim 0$). BOOTES, despite its lower number density, shows a modest satellite fraction, $f_{\rm sat} \sim 0.15$, and slightly higher $M_{\rm eff}$ and $b_{\rm eff}$. This suggests that the COSMOS field contains a larger number of moderate-mass halos, each hosting a single central ULIRG, whereas BOOTES is populated by fewer but more massive halos that are capable of hosting multiple ULIRGs as satellites. Given the broad uncertainty on \(f_{\rm sat}\), especially in COSMOS, this interpretation should be regarded as suggestive rather than definitive.

\section{CONCLUSION AND DISCUSSION}
\begin{figure}
\plotone{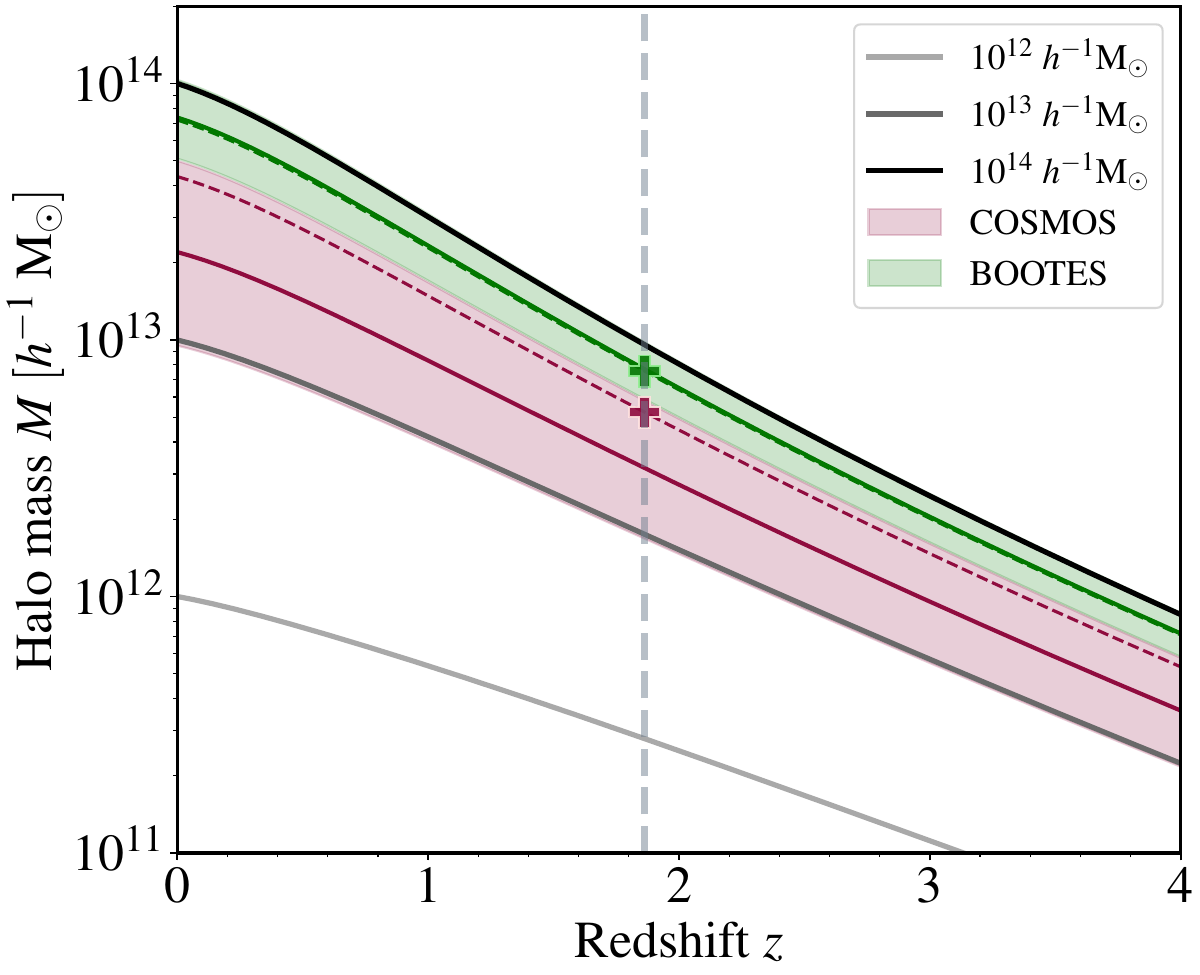}
\caption{Halo mass assembly history as a function of redshift by \citet{BehrooziEtal2013}. Colored solid lines are evolution trajectories of the {\tt Posterior PDF} $M_{\rm eff}$ of ULIRGs at $z=1.86$, and corresponding uncertainties in $M_{\rm eff}$ are shown as shaded regions. At $z=0$, the halo masses of these ULIRGs would reach to $10^{13.34}\ \msun$ and $10^{13.86}\ \msun$ respectively. For comparison, the {\tt Best fit} $M_{\rm eff}$ are also presented as plus symbols, and their evolution trajectories as dashed color lines.}
\label{fig:halo_mass}
\end{figure}

In this study, we have determined the host halo masses of $z\sim 2$ ULIRGs in the COSMOS and BOOTES fields via angular clustering analysis. Both the linear bias model and the HOD framework yield consistent characteristic masses, with $M_{\rm h}\sim 10^{12.7}\ \msun$; the HOD provides tighter constraints by incorporating the one-halo term, giving $M_{\rm eff}=10^{12.50}\ \msun$ and $10^{12.89}\ \msun$ respectively.  Using the halo mass evolution model of \citet{BehrooziEtal2013}, we find that ULIRGs at $z\sim 2$ are expected to evolve into halos of mass $10^{13.34}\ \msun$ and $10^{13.86}\ \msun$ respectively at $z\sim 0$ (Fig.~\ref{fig:halo_mass}).

A notable result is the possible difference in satellite occupation between the two fields with different number densities. COSMOS ULIRGs are consistent with being pure central galaxies ($f_{\rm sat}\sim 0$, 68\% upper bound $\sim 0.20$), whereas BOOTES shows a modest but non-negligible satellite fraction $f_{\rm sat}\sim 0.15$. This difference is noteworthy because COSMOS has the higher number density ($457\,{\rm deg}^{-2}$ versus $270\,{\rm deg}^{-2}$).

Our effective halo mass measurements are in agreement with the results of \citet{StarikovaEtal2012}, but significantly lower than that reported by \citet{FarrahEtal2006}, even after correcting for the $h^{-1}$ factor in the mass unit. While \citet{MagliocchettiEtal2008} did not provide an effective host halo mass but reported a bias parameter $b=6.17$ for their $z\sim 2$ sample, which is significantly higher than our measurement. Furthermore, their $\log M_{\rm min}\approx 12.8$ is apparently lower than what we have obtained here. The origin of these discrepancies is unclear, but they may reflect differences in sample construction, parameter estimation methods, or a manifestation of cosmic variance among different fields. In comparison, our results are broadly consistent with those reported for SMGs at similar redshifts \citep[e.g.,][]{HickoxEtal2012, ChenEtal2016, WilkinsonEtal2017, AmvrosiadisEtal2019, LimEtal2020, StachEtal2021}. Although more flexible HOD models have been developed to describe star-forming galaxies \citep[e.g.,][]{GeachEtal2012,OkumuraEtal2021,OsatoEtal2023}, we adopt the commonly used Zheng model because the additional freedom in those models would lead to poorly constrained parameters and stronger degeneracies without a clear gain in physical insight for the present data.

We acknowledge that the satellite fraction in COSMOS is only weakly constrained, the radial distribution function of galaxies might vary across different fields, and that the Gaussian likelihood adopted for the ACF (Eq.~5) is an approximation; however, given the signal‑to‑noise ratio of the current ULIRG samples, these choices are adequate and do not affect our main conclusions. The jackknife covariance matrix (Appendix~\ref{sec:jk_num}) yields stable parameter estimates, and we have verified that alternative treatments of the integral constraint change the results by less than $2\%$.

Finally, recent JWST observations have revealed that $z\sim2$ ULIRGs form a diverse population, with multiple evolutionary pathways \citep[]{HuangEtal2023}. The field‑to‑field variation in satellite occupation seen here may be a manifestation of this diversity as part of cosmic variance. Combining current archival data (e.g., from Spitzer/MIPS) with future larger‑area surveys (e.g., with \textit{Euclid} or \textit{Roman}) will provide sufficient statistics to constrain the HOD parameters more precisely and to test whether the environmental dependence of ULIRG multiplicity is a universal feature of galaxy evolution at cosmic noon.

\begin{acknowledgements}
JP appreciates funding from NSFC of grant No.12273049. ML acknowledges the support from the National Key Research and Development Program of China (No.2022YFA1602903) and the National SKA Program of China (Nos. 2022SKA0110201 and 2022SKA0110200). CC is supported by the Chinese Academy of Sciences South America Center for Astronomy (CASSACA) Key Research Project E52H540101 and E52H540301. This work is also jointly sponsored by the China Manned Space Program with grant No.CMS-CSST-2025-A07, and the Chinese Academy of Sciences (CAS) through a grant to the CAS South America Center for Astronomy.
\end{acknowledgements}

\appendix
\section{Technical Details of the 2PCF and Covariance Matrix Estimation}
\subsection{Covariance matrix}
\label{sec:jk_num}
\begin{figure*}
\plottwo{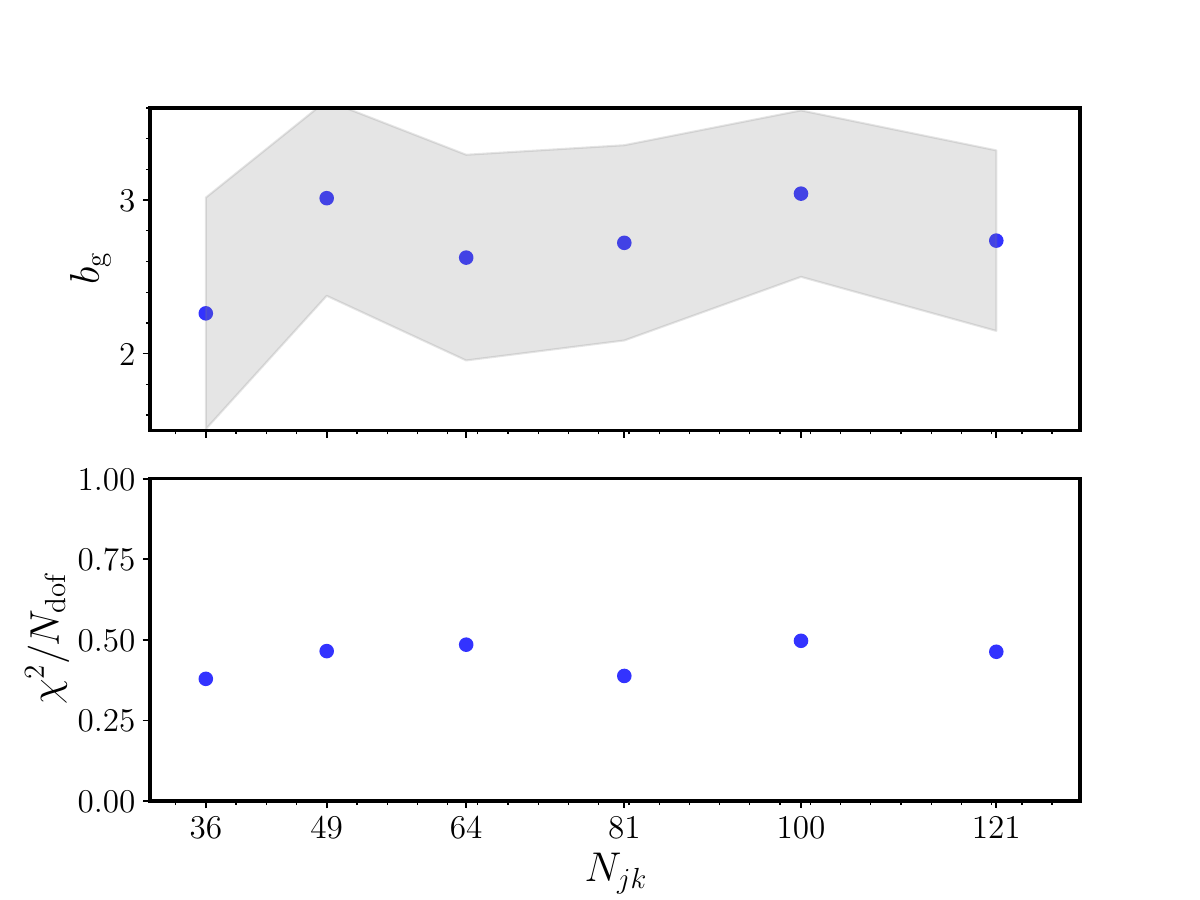}{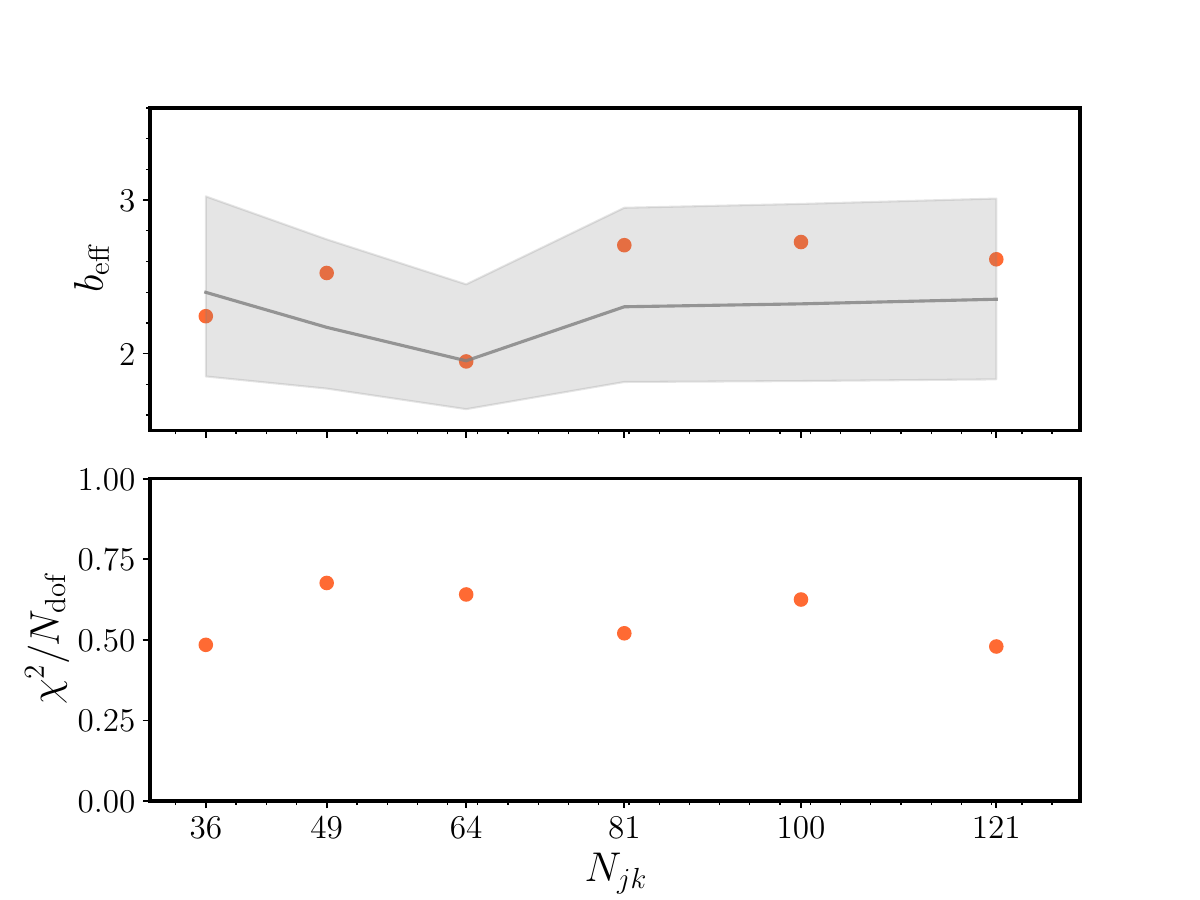}
\caption{The MLE analysis with linear model (left panel) and HOD parameter sampling (right panel) in COSMOS, using covariance matrix estimated with jackknife realizations $N_{jk}=36,49,64,81,100,121$. The upper and lower subpanels in each panel show the bias parameter and the corresponding reduced $\chi^2$.}
\label{fig:bg_chi2}
\end{figure*}

Galaxy samples in the COSMOS and BOOTES fields span sky areas of approximately $87.6^\prime \times 89.4^\prime$ and $216.2^\prime \times 215.8^\prime$, respectively. When constructing jackknife subsamples, the survey area is typically divided into regions as small as $\sim 1/10$ of the side length, yielding $N_{jk}\sim 100$. However, to robustly determine the optimal number of subsamples for our covariance matrices, we conducted a series of numerical experiments.

For the COSMOS field, we generated jackknife subsamples with $N_{\rm jk}=36, 49, 64, 81, 100$, and $121$, and estimated the covariance matrix using Eq.~\ref{eq:covmat}. We then performed $\chi^2$ minimization for the linear model and MCMC sampling for the HOD, as presented in Sections~\ref{sec:linmodel} and \ref{sec:hodmodel}. Figure~\ref{fig:bg_chi2} shows the derived linear bias $b_{\rm g}$ and the HOD-inferred effective bias $b_{\rm eff}$ as functions of $N_{\rm jk}$. The linear bias stabilizes once $N_{\rm jk} \gtrsim 36$, while $b_{\rm eff}$ is robust across the full tested range, exhibiting only modest fluctuations at $N_{\rm jk}=36$ and $64$. The reduced $\chi^2$ follows a similar trend for both modeling approaches. We adopt $N_{\rm jk}=100$ as our fiducial choice: it sits well within the stable plateau of inferred parameters, while keeping the jackknife subregions sufficiently large to avoid overfitting. We use this same $N_{\rm jk}$ for both the linear and HOD analyses to ensure consistency. The HOD reduced $\chi^2$ varies only weakly between $N_{\rm jk}=49$ and $100$, so our choice does not compromise the fit quality.

For the BOOTES sample, we performed a similar numerical experiment with $N_{\rm jk}=64, 81, 100$, and $121$. The inferred $b_{\rm g}$, $b_{\rm eff}$, and reduced $\chi^2$ values vary only weakly over this range. We adopt $N_{\rm jk}=81$ as our fiducial choice, which yields a reliable covariance estimate without overfitting. Adopting $N_{\rm jk}=100$ or $121$ would not alter the inferred parameters beyond their quoted uncertainties.

\subsection{Integral constraint}
\label{sec:ic_test}
To measure the effects of the integral constraint with Eq.~\ref{eq:ic}, it is necessary to have $RR$ up to the largest scale allowed by the sample and a theoretical model to facilitate the calculation \citep{RocheEales1999}. \citet{GeachEtal2012} argues that the scaled nonlinear 2PCF of dark matter shall be sufficient, since the weights $RR(\theta_i)/\sum_i RR(\theta_i)$ at small scales are much less than those in the linear regime. We have experimented with three different models of $\omega(\theta)$: 
\begin{itemize}
\item $b_{\rm g}^2 \omega_{\rm l}$ ($\omega_{\rm l}$ is the linear A2PCF of dark matter);
\item $b_{\rm g}^2 \omega_{\rm nl}$ ($\omega_{\rm nl}$ is the nonlinear A2PCF of dark matter);
\item a hybrid function that reads $A\theta^{-\gamma}$ on $\theta < 1.3^\prime$, while at large scales it switches to $b_{\rm g}^2 \omega_{\rm l}$. 
\end{itemize}
The numerical results of our COSMOS and BOOTES samples demonstrate that the values of $\Delta\omega$ obtained by these different models differ by $< 2\%$.

\bibliographystyle{aasjournal}
\bibliography{acos} 

\end{document}